\documentclass[aps,showpacs,twocolumn,pre,superscriptaddress]{revtex4}

\usepackage{graphicx}
\usepackage{amsmath}
\usepackage{amssymb}

\begin{document}

\title{Diffusive Counter Dispersion of Mass in Bubbly Media}

\author{Denis S.\ Goldobin}
\affiliation{Department of Mathematics, University of Leicester,
             Leicester LE1 7RH, United Kingdom}
\affiliation{Institute of Continuous Media Mechanics, UB RAS,
             Perm 614013, Russia}
\author{Nikolai V.\ Brilliantov}
\affiliation{Department of Mathematics, University of Leicester,
             Leicester LE1 7RH, United Kingdom}

\begin{abstract}
We consider a liquid bearing gas bubbles in a porous medium. When
gas bubbles are immovably trapped in a porous matrix by
surface-tension forces, the dominant mechanism of transfer of gas
mass becomes the diffusion of gas molecules through the liquid.
Essentially, the gas solution is in local thermodynamic
equilibrium with vapor phase all over the system, i.e., the solute
concentration equals the solubility. When temperature and/or
pressure gradients are applied, diffusion fluxes appear and these
fluxes are faithfully determined by the temperature and pressure
fields, not by the local solute concentration, which is enslaved
by the former. We derive the equations governing such systems,
accounting for thermodiffusion and gravitational segregation
effects which are shown not to be neglected for geological
systems---marine sediments, terrestrial aquifers, etc. The results
are applied for the treatment of non-high-pressure systems and
real geological systems bearing methane or carbon dioxide, where
we find a potential possibility of the formation of gaseous
horizons deep below a porous medium surface. The reported effects
are of particular importance for natural methane hydrate deposits
and the problem of burial of industrial production of carbon
dioxide in deep aquifers.
\end{abstract}


\maketitle

\section{Introduction}
The diffusion of solute gases in liquids is a well-studied and
relatively well-understood problem; see,
e.g.~\cite{Hirschfelder-Curtiss-Bird-1954,Bird-Stewart-Lightfoot-2007}.
In the classical formulation of the problem, the diffusion flux of
guest atoms or molecules is expressed in terms of various
thermodynamic ``forces''---concentration, pressure, and
temperature gradients. Usually, the liquid-gas interface does not
influence the nature of the bulk diffusion and determines only the
boundary conditions for the flux. There exists, however, a vast
class of systems, called ``bubbly liquids'',
e.g.~\cite{Yurkovetsky-Brady-1996}, where the liquid-gas interface
maintains the gas-solution saturation all over the liquid volume
and serves as a source/sink for the diffusion flux of the solute
molecules (i.e., the diffusion flux does not change the solute
concentration with time, but redistributes the mass between gas
bubbles). Thus the liquid-gas interface plays an important role in
the bulk diffusion of solutes. On the macroscopic scale, the
diffusive mass transport of solutes may be very unusual and
surprising in these systems.

\begin{figure}[!tb]
\center{
 \includegraphics[width=0.26\textwidth]%
 {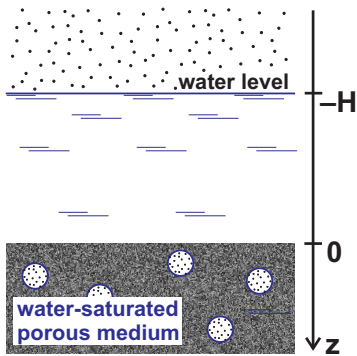}
}

  \caption{Typical system of bubbly liquid filling the porous medium.
The pressure, temperature, and parameters of the porous matrix may
vary vertically, but are constant horizontally. The figure, with
water above the bubble-rich sediments,  sketches numerous
geological systems.}
  \label{fig1}
\end{figure}

In the present study, we address bubbly liquids with immobile
bubbles. These may be trapped by the surface forces in a porous
matrix, therefore, for concreteness, we will consider liquids in
porous media. Among numerous examples of such systems, which are
of great practical importance, are the oil-bearing porous massifs,
seabeds in organic carbon-rich marine
sediments~\cite{Archer-2007,Davie-Buffett-2001,Davie-Buffett-2003,Garg_etal-2008,Haacke-Westbrook-Riley-2008},
aquifers~\cite{Donaldson-etal-1997,Donaldson-etal-1998}, peat-bogs
and swamps. For these systems, the presence of immobilized gas
bubbles or liquid drops is well established experimentally. The
respective solute gases include methane, carbon dioxide, oxygen,
nitrogen, etc. Moreover, the problem of diffusion of the carbon
dioxide in porous medium, filled with bubbly liquid (water) is
directly related to the problem of burial of industrial release of
$\mathrm{CO_2}$, e.g.~\cite{Holloway-1997,Rochelle_etal-2009}. In
Fig.~\ref{fig1}, we sketch a typical system, where the diffusion
of a solute gas takes place in a bubbly liquid in a porous matrix.

We will study diffusion transport on large space and time scales.
Namely, we will be interested in the space scale much larger than
the characteristic size of a gas bubble and in the time scale much
larger than the characteristic relaxation time for gas dissolution
in liquid; hence we assume that all over the volume, the solute
gas is saturated in the solution. Indeed, for marine sediments the
solution relaxation times measured by hours are small compared to
the global evolution time scales, which can be as large as
millions of years. To estimate the pore size $l$ needed to trap a
bubble of the comparable size we notice that the surface-tension
forces, equal to $\sigma l$ ($\sigma$ is the surface tension),
should overwhelm the buoyancy force $\rho_\mathrm{liq}gl^3$
($\rho_\mathrm{liq}$ is the density of a liquid; $g$ is the
gravity). This yields the estimate,
$l<\sqrt{\sigma/\rho_\mathrm{liq}g}$, implying the maximal pore
size $l < 2.7\,{\rm mm}$ for water, which suggests trapping even
for sands. We consider the immobilization of pore-size bubbles,
because a big moving bubble is unstable to
splitting~\cite{Lyubimov-etal-2009}; it either stops or
experiences splitting into smaller bubbles until their size
becomes comparable to the pore size.
 We assume that our space scale is much
larger than $l$. It is not straightforward to estimate the
relaxation time for the gas dissolution; therefore, we just assume
that the condition of the gas saturation in liquid is always
fulfilled.

Another important feature of bubbly liquids in porous media
(relevant for many geological systems) is the presence of a
temperature gradient (usually due to the geothermal gradient) and
a pressure gradient (due to the hydrostatic pressure). On the
microscopic level, the temperature gradient causes the
thermodiffusion (the so-called ``Soret
effect''~\cite{Soret-1879,Jones-Furry-1946,Hirschfelder-Curtiss-Bird-1954,Bird-Stewart-Lightfoot-2007}),
while the pressure gradient manifests the presence of the
gravitational forces. Due to different mobilities and masses of
the solute and solvent molecules these may cause an additional
mass flux. Although the importance of thermodiffusion for soil gas
exchange has been reported in Ref.~\cite{Richter-1972}, only the
diffusion flux due to the concentration gradient (Fickian
diffusion) has been investigated so
far~\cite{Davie-Buffett-2001,Davie-Buffett-2003,Garg_etal-2008,Haacke-Westbrook-Riley-2008}.

In the present study, we consider the diffusive transport of a
solute gas in liquid in a porous medium without any significant
through-flow in the system; that is, we consider only molecular
and not hydrodynamic
diffusion~\cite{Donaldson-etal-1997,Donaldson-etal-1998,Saffman-1959,Sahimi-1993}.
We show that for large space and time scales, which assume the
saturation condition for the solute gas, and under temperature and
pressure gradients a novel and interesting phenomenon may be
observed---the diffusive accumulation of gas in certain regions,
caused by the non-Fickian transport of solutes.

The paper is organized as follows. In Sec.~\ref{sec:DBL} the
diffusion flux of a solute gas in bubbly liquid is considered in
the presence of pressure and temperature gradients; the evolution
equation for the gas concentration is obtained. In
Sec.~\ref{sec:Geo} we demonstrate the existence of a novel
phenomenon---the diffusive accumulation of solute gas in certain
regions. We analyze this effect for real  geological systems---the
methane or carbon dioxide bearing massifs in the presence of the
geothermal gradient. Finally, in  Sec.~\ref{sec:Concl}, we
summarize our findings. Some technical details are given in the
Appendixes.

%%%%%%%%%%%%%%%%%%%%%%%%%%%%%%%%%%%%%%%%%%%%%%%%%%%%%%%%%%%%%%%
%%%%%%%%%%%%%%%%%%%%%%%%%%%%%%%%%%%%%%%%%%%%%%%%%%%%%%%%%%%%%%%

\section{Diffusion transport in bubbly liquid with gradients of
         thermodynamic properties}\label{sec:DBL}
The diffusion current of guest particles in a host material  in
the presence of concentration and temperature gradients and an
external potential generally has three components, each being the
response to their respective driving force: The Fickian flux, due
to the concentration gradient, the thermodiffusion flux, due to
the temperature gradient, and the flux due to the mobility of
particles subjected to an external force. When there are no
external forces except gravity, which also creates the pressure
gradient, the diffusion flux of a dilute solution
reads~\cite{Bird-Stewart-Lightfoot-2007}
\begin{equation}
\vec{J}=-DX\left[\frac{\nabla X}{X}
 +\alpha_T\frac{\nabla T}{T}-\frac{\tilde{M}\vec{g}}{RT}\right],
\label{eq-DBL-01}
\end{equation}
where $X$ is the molar fraction (concentration) of the solute in
the solution, $D$ is the molecular diffusion coefficient of the
solute molecules in the solvent, $\alpha_T$ is the thermodiffusion
constant ($\alpha_T/T$ is the Soret coefficient), and
 $R=8.31{\rm J\,K^{-1}mol^{-1}}$
is the universal gas constant.
$\tilde{M}=M^\mathrm{g}-N_1M^\mathrm{host}$, where $M^\mathrm{g}$
and $M^\mathrm{host}$ are the molar masses of solute and solvent
molecules, and $N_1$ is the number of solvent molecules in the
volume occupied by one solute molecule in the solution. For
specific gases we derive $N_1$ from experimental data in
Appendix~\ref{AppA}.

When the liquid is saturated with gas bubbles and the bubbles are
in local thermodynamic equilibrium with the solution, the
concentration of the solute in the solvent equals solubility,
$X=X^{(0)}$, everywhere in the liquid. The thermodynamic
equilibrium implies the equality of the chemical potentials of the
gas dissolved in the liquid ($\mu_\mathrm{liq}^\mathrm{g}$) and
that of the vapor phase ($\mu_\mathrm{vap}^\mathrm{g}$), that is,
$\mu_\mathrm{liq}^\mathrm{g}=\mu_\mathrm{vap}^\mathrm{g}$.
According to the thermodynamic laws, the chemical potential of the
vapor phase depends exclusively on temperature $T$ and pressure
$P$,
$\mu_\mathrm{vap}^\mathrm{g}=\mu_\mathrm{vap}^\mathrm{g}(T,P)$.
Hence the solute concentration $X$ is not a free variable, but a
function of the local temperature and pressure,
$X(\vec{r})=X^{(0)}\left(T(\vec{r}),P(\vec{r}) \right) $; the same
holds true for the solute flux $\vec{J}(\vec{r})$. In
Appendix~\ref{AppB} of this paper we provide the calculation of
high-pressure aqueous solubility of gases based on a scaled
particle theory~\cite{Pierotti-1976} amended with implementation
of van der Waals' equation of state for the vapor phase.

In the present study we consider the problem of diffusion
transport in a bubbly liquid in the context of real geological
systems, where pressure varies from one to a few hundred
atmospheres (this refers to a few kilometers deep water column,
see Fig.~\ref{fig1}). To illustrate the approach, we will discuss
first a more simple case of moderate pressures $P<100\,{\rm atm}$,
which allows an explicit analytical treatment. The results for the
general case, obtained numerically, will be also shown.

According to Eq.~(\ref{eq-ScP-05}) in Appendix~\ref{AppB}, for
moderate pressures and far from the solvent boiling temperature
the solubility depends on $T$ and $P$ as follows
$$
X^{(0)}(T,P) \simeq X^{(0)}(T_0,P_0)\frac{T_0}{T}\frac{P}{P_0}
 \exp\left[q\left(\frac{1}{T}-\frac{1}{T_0}\right)\right] \, ,
$$
where $T_0$ and $P_0$ are reference values, the choice of which is
guided merely by the convenience reason, and $X^{(0)}(T_0,P_0)$ is
the solubility at the reference temperature and pressure; the
parameter $q \equiv -G_i/k_\mathrm{B}$ is given in
Tab.~\ref{params}. Pressure is moderate in a sense that one can
neglect van der Waals' effects and the influence of pressure on
the Gibbs free energy of the cavity formation for the guest
molecule in the solvent. The latter is represented by the term
$(-BP/T)$ in the exponential of Eq.~(\ref{eq-ScP-05}), where
constant $B\approx10^{-6}\,{\rm K/Pa}$ (see Tab.~\ref{params}),
and becomes significant only for pressures as high as several
hundreds atmospheres. The limiting formula is free of the molar
fraction $Y$ [Eq.~(\ref{eq-ScP-06})] of the solvent molecules in
gas bubbles because it is assumed to be zero. This assumption is
valid several tens of Kelvins below the water boiling temperature,
while the systems where liquids are at near-boiling conditions,
say geysers, are beyond the scope of the present study.

Using
 $\nabla X^{(0)}=X^{(0)}[P^{-1}\nabla P-(1+q/T)T^{-1}\nabla T]$,
one obtains for the total flux:
\begin{equation}
\vec{J}\simeq -DX^{(0)}\left[\frac{\nabla P}{P}
 -\left(1-\alpha_T+\frac{q}{T}\right)\frac{\nabla T}{T}
 -\frac{\tilde{M}\vec{g}}{RT}\right].
\label{eq-DBL-02}
\end{equation}
Note that the flux does not depend on the solute concentration or
its gradient as independent fields.

Generally, the flux (\ref{eq-DBL-02}) possesses a nonzero
divergency $\nabla\cdot\vec{J}$, which implies the existence of
sources and sinks for the solute mass. Obviously, the gas  bubbles
serve as the mass reservoir, which provides the respective sources
and sinks. Hence the mass balance reads
\begin{equation}
\frac{\partial X_\mathrm{b}}{\partial t}+\nabla\cdot\vec{J}=0 \, ,
 \label{eq-DBL-03}
\end{equation}
where $X_\mathrm{b}$ is the molar fraction of bubbles in the
bubbly fluid, that is, in the system comprised by bubbles and
liquid. For the space scales addressed here we do not need to
consider the microscopic processes of the bubbles growth and
clustering as in~\cite{Dominguez-Bories-Prat-2000}, and the single
characteristic, $X_\mathrm{b}(\vec{r})$, suffices in our case.

Hence for moderate pressures we obtain the following equation for
the evolution of a nondissolved gas phase:
\begin{equation}
\frac{\partial X_\mathrm{b}}{\partial t} \simeq \nabla\cdot\left(
 DX^{(0)}\left[\frac{\nabla P}{P}
 -\left(1-\alpha_T+\frac{q}{T}\right)\frac{\nabla T}{T}
 -\frac{\tilde{M}\vec{g}}{RT}\right]
\right).
\label{eq-DBL-04}
\end{equation}
Eqs.~(\ref{eq-DBL-03}) and (\ref{eq-DBL-04}) are written for the
case when the bubbles occupy a small fraction of the fluid volume,
which is typical for most geological systems,
e.g.~\cite{Davie-Buffett-2001,Davie-Buffett-2003}. Moreover, the
corrections needed to account for the finite fraction of the
bubble volume are always quantitative and never qualitative.
Indeed, the direction of the solute flux is merely determined by
the factor in the square brackets in the right-hand side of
Eq.~(\ref{eq-DBL-04}), which does not depend on the bubble volume.

\section{The diffusive accumulation and non-Fickian flux of solutes}
\label{sec:Geo}

In what follows we consider the one-dimensional diffusion of
solutes in bubbly liquids in the presence of temperature and
density gradients, using for the latter quantities some typical
geological values. The model of one-dimensional diffusion is
motivated by the fact that real geological systems are much more
uniform along two directions (say horizontal) than along the third
direction (say vertical). Then we have a laterally uniform system
with vertical gradients of its thermodynamic properties. For the
sake of definiteness we analyze  the case of seabed sediments,
Fig.~\ref{fig1}, where the depth below the seafloor is measured by
the $z$ coordinate. Then we have $z$-dependent hydrostatic
pressure and $z$-dependent temperature due to the geothermal
temperature gradient:
$$
 P(z)=P_0+\rho_\mathrm{liq}g(z+H),\qquad
 T(z)=T_\mathrm{sf}+Gz.
$$
Here $P_0$ is the atmospheric pressure, $H$ is the height of the
pure-water layer above the bubble-bearing porous medium,
$T_\mathrm{sf}$ is the temperature of the seafloor and  $G$ is the
geothermal gradient. Nonlinearity of the temperature
profile~\cite{Goldobin-EPL-2011} is neglected in our study because
it is system specific and not a principal ingredient for the
phenomenon we consider. The assumption of a linear temperature
profile is typical for studies on physical processes in marine
sediments~\cite{Davie-Buffett-2001,Davie-Buffett-2003}.

%%%%%%%%%%%%%%%%%%%%%%%%%%%%%%%%%%%%%%%%%%%%%%%%%%%%%%%%%%%%%%%
\begin{figure}[!tb]
\center{
 \includegraphics[width=0.48\textwidth]%
 {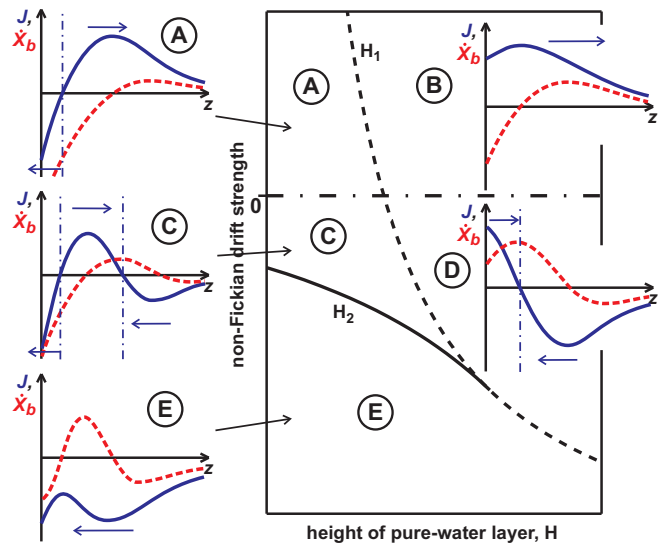}
}

  \caption{(Color online) Diagram of diffusive regimes
on the plane $\beta$--$H$ ($\beta$ quantifies the non-Fickian
drift strength) for moderate pressures. The critical heights $H_1$
and $H_2$ are given by Eqs.~(\ref{eq-Geo-02}) and
(\ref{eq-Geo-03}), respectively. In the plug-in plots the
dependence of the solute flux $J$ (blue solid lines) and of the
bubble growth rate $\dot{X}_b$ (red dashed lines) on the depth $z$
below the seafloor is shown.}
  \label{fig2}
\end{figure}
%%%%%%%%%%%%%%%%%%%%%%%%%%%%%%%%%%%%%%%%%%%%%%%%%%%%%%%%%%%%%%%

%%%%%%%%%%%%%%%%%%%%%%%%%%%%%%%%%%%%%%%%%%%%%%%%%%%%%%%%%%%%%%%
\begin{figure*}[t]
\center{
 {\sf (a)\hspace{-2mm}}
 \includegraphics[width=0.82\textwidth]%
 {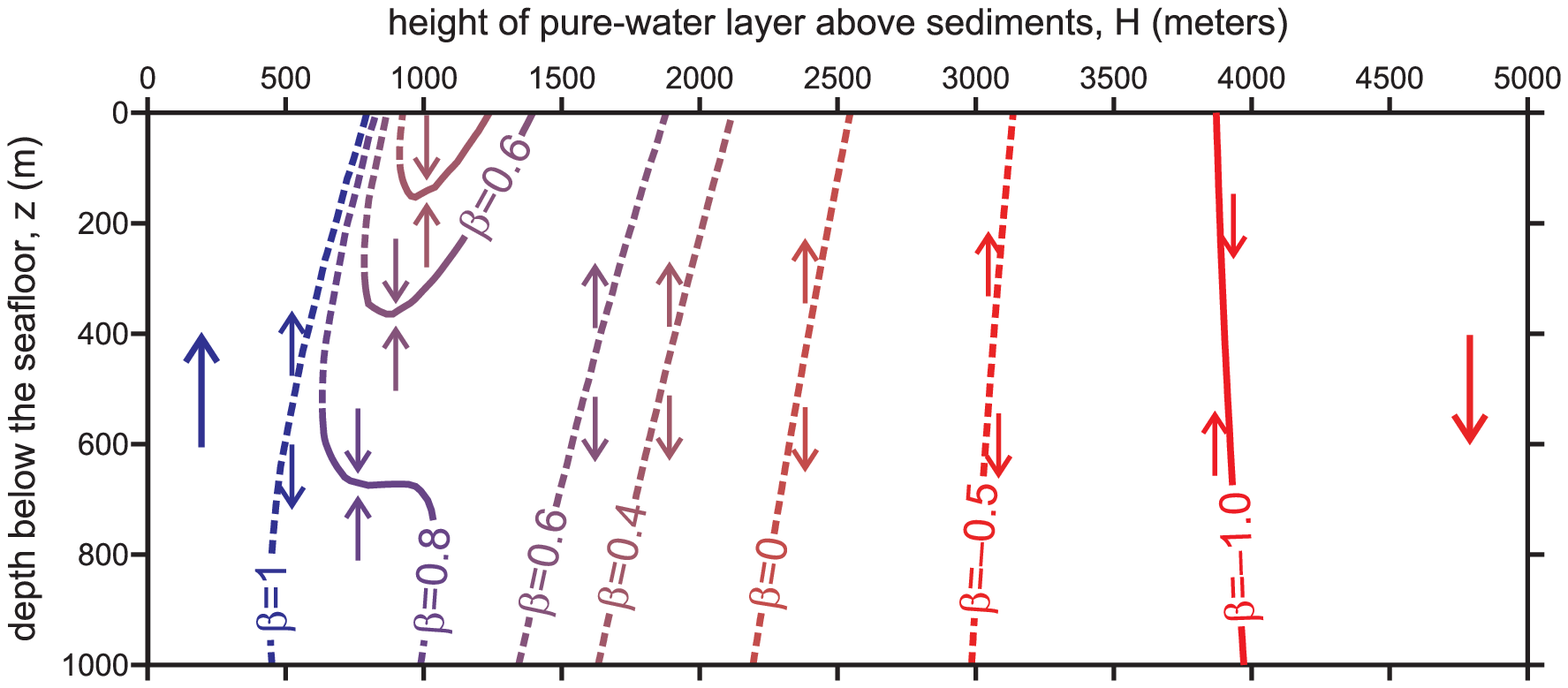}
 \\[8pt]
 {\sf (b)\hspace{-2mm}}
 \includegraphics[width=0.82\textwidth]%
 {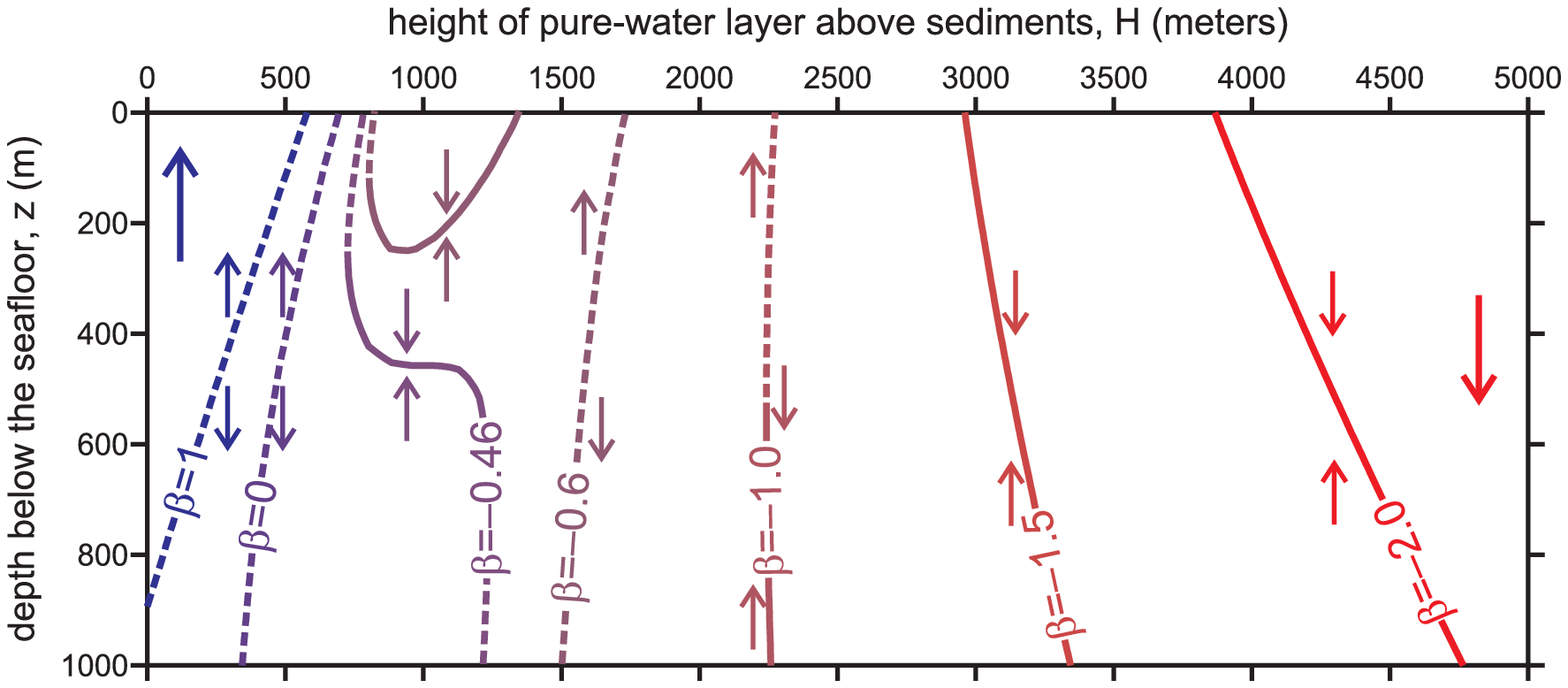}
}
  \caption{(Color online) Direction of the mass flux and
the position of the flux inversion point as the function of the
height of pure-water column above the sediments. Different curves
correspond to different values of the non-Fickian parameter
$\beta$ for methane. Calculations have been performed for the
seafloor temperature $T_\mathrm{sf}=277{\rm K}$ and geothermal
gradient $G=40{\rm K/km}$~(a) and $G=60{\rm K/km}$~(b) for the
general case, without the approximation of moderate pressures. The
flux inversion points with negative divergency (accumulating the
solute gas) are plotted with the solid curves, the points with
positive divergency (``repelling'' the solute gas) are shown with
the dashed curves.}
  \label{fig3}
\end{figure*}
%%%%%%%%%%%%%%%%%%%%%%%%%%%%%%%%%%%%%%%%%%%%%%%%%%%%%%%%%%%%%%%

%%%%%%%%%%%%%%%%%%%%%%%%%%%%%%%%%%%%%%%%%%%%%%%%%%%%%%%%%%%%%%%
\begin{figure}[t]
\center{
 \includegraphics[width=0.36\textwidth]%
 {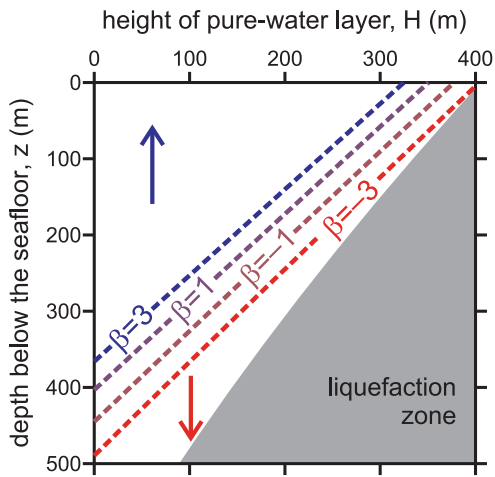}
}
  \caption{(Color online) Direction of the mass flux and
the position of the flux inversion point for carbon dioxide.
Calculations have been performed for the seafloor temperature
$T_\mathrm{sf}=277\,{\rm K}$ and geothermal gradient $G=40\,{\rm
K/km}$ for the general case, that is, without the the
approximation of moderate pressures. For details see the caption
for Fig.\,\ref{fig3}.}
  \label{fig4}
\end{figure}
%%%%%%%%%%%%%%%%%%%%%%%%%%%%%%%%%%%%%%%%%%%%%%%%%%%%%%%%%%%%%%%

Using the above expressions for the  hydrostatic pressure and
geothermal gradient, we obtain for the diffusion flux
(\ref{eq-DBL-02}) for moderate pressures:
\begin{eqnarray}
J \simeq DX^{(0)}\left[-\frac{1}{z+H+P_0/\rho_\mathrm{liq}g}
\right.\qquad\qquad
\nonumber\\
\left.
 +\left(1+\beta
 +\frac{q}{T_\mathrm{sf}+Gz}\right)\frac{1}{z+T_\mathrm{sf}/G}
 \right],
\label{eq-Geo-01}
\end{eqnarray}
where
\[
\beta \equiv -\alpha_T+\frac{\tilde{M}g}{RG}.
\]
Eq.~(\ref{eq-Geo-01}) with $\beta=0$ corresponds to the Fickian
diffusion, $\vec{J}=-D\nabla{X}$, while the coefficient $\beta$
quantifies the strength of the non-Fickian contribution to
diffusion flux. Let us consider the consequences of the above
model for the typical geological systems, like marine sediments.
Fig.~\ref{fig2} illustrates  the variety of possible diffusive
regimes. For the gases with {\em positive} $\beta$, the gas leaves
the sediments. In shallow seas the gas  diffuses upwards, to the
sea, in the thin upper layer of sediments and downwards below this
layer [regime (A) in Fig.~\ref{fig2}]. When the sea depth $H$
exceeds a certain value $H_1$, the upper layer disappears and the
gas diffuses downwards all over the seabed [regime (B)]; for {\it
moderate pressures} the threshold depth reads,
\begin{equation}
H_1=\frac{T_\mathrm{sf}}{G\left(\displaystyle
 1+\beta+q/T_\mathrm{sf}\right)}
 -\frac{P_0}{\rho_\mathrm{liq}g}.
\label{eq-Geo-02}
\end{equation}
For gases with {\em negative} $\beta$, the direction of the flux
very deep below the seafloor turns upwards. In deep seas, $H>H_1$
[regime (D)], the solute  accumulation zone appears: The solute
gas from the upper and deep layers of sediments diffuses to this
zone. For shallow seas, $H<H_1$ [regime (C)], the solute gas
accumulation zone loses its contact with the seafloor. In this
case the solute gas from the region just beneath the seafloor
migrates upwards, to the sea.

Thus we reveal  a novel and surprising phenomenon---the formation
of a gas accumulation zone: Instead of the tendency of the gas to
spread uniformly over the system, as expected for the common
diffusion, the gas concentrates in narrow zones, due to the
non-Fickian diffusion against the direction dictated by the
concentration gradient.

Finally, if the positive thermodiffusion is strong enough, one
observes the regime (E), where no gas-retracting zone exists and
all the solute migrates upwards, to the sea. Nevertheless, the
inhomogeneity of the diffusion flux results in the bubble growth
zone. For {\it moderate} pressures, the depth $H$ of the water
body beneath which this diffusion regime occurs is bounded from
above by
\begin{eqnarray}
&&
 H_2=\frac{T_\mathrm{sf}}{G}-\frac{P_0}{\rho_\mathrm{liq}g}
 -\frac{q}{G\left(1+\sqrt{-\beta}\right)^2}
\label{eq-Geo-03}\\
&&
 \mbox{ for }\beta\in\left[
 -\frac{1}{4}\left(\sqrt{1+\frac{4q}{T_\mathrm{sf}}}-1\right)^2,
\right.
\nonumber\\
&&\qquad\qquad\qquad
\left.
 -\left(\sqrt{\frac{q}{T_\mathrm{sf}-P_0G/\rho_\mathrm{liq}g}}-1\right)^2\right].
\nonumber
\end{eqnarray}

For high pressure, the solubility and diffusion flux are more
gas-specific and each particular case requires a separate
analysis. In Figs.~\ref{fig3} and \ref{fig4}, the results for
methane and carbon dioxide are shown for the general case, where
Eqs.~(\ref{eq-DBL-01}), (\ref{eq-ScP-05}), and (\ref{eq-ScP-06})
have been used without the simplifying approximations adopted for
the moderate pressures. The range of parameters used in
Figs.~\ref{fig3} and \ref{fig4} is chosen to be relevant for
practical applications: The range of pressures for methane
corresponds to the possible location of the gas-hydrate deposits
(methane is one of the components of the hydrate),  while the
pressures for carbon dioxide may be essential for the process of
burial of industrial $\mathrm{CO_2}$ in deep aquifers.

For methane (Fig.~\ref{fig3}), we consider the upper $1-{\rm km}$
layer of sediments for sea depths ranging from $0\,{\rm km}$ to
$5\,{\rm km}$ with the seafloor temperature
$T_\mathrm{sf}=277\,{\rm K}$ (which corresponds to maximal water
density at atmospheric pressure). For a small geothermal gradient,
$G=20\,{\rm K/km}$, methane diffuses upwards, to the sea, for any
value of the non-Fickian parameter $\beta$ and any sea depth $H$.
For a larger $G$, the transport significantly depends on $\beta$
and $H$ (Fig.~\ref{fig3}). For the typical value of $G=\,40{\rm
K/km}$
 \cite{Davie-Buffett-2001,Davie-Buffett-2003}
methane diffuses upwards in shallow seas and downwards for deep
ones; the sea depth $H$ for which the flux at $z=0$ reverses
increases as $\beta$ decreases. Remarkably, for the range of the
non-Fickian parameter $0.3\lesssim\beta\lesssim 0.9$, the gas
accumulation zone (which blocks the release of methane into the
sea) is located in the seabed at the depth ranging from $0.7\,{\rm
km}$ to $1.6\,{\rm km}$ (solid lines in Fig.~\ref{fig3}a). For
$\beta\lesssim-1$, this zone appears in sediments under deep water
bodies, $H\gtrsim3.9\,{\rm km}$. In Fig.~\ref{fig3}b, where
$G=60\,{\rm K/km}$, one can see how the diffusion regimes are
affected by further increase of the geothermal gradient. We wish
to stress, that the reported phenomenon originates exclusively due
to the non-Fickian diffusion. The oversimplified models based on
the Fickian law with $\beta=0$,
e.g.~\cite{Davie-Buffett-2001,Davie-Buffett-2003}, cannot describe
the novel effect of the formation of gas accumulation zone.

The main difficulty of the practical application of our theory is
the lack of precise data for the thermodiffusion constant, that
is, the lack of a trustworthy value for the coefficient $\beta$.
For instance, the authors are not aware of any experimental data
on the thermodiffusion of methane in water. Theoretical studies,
e.g.~\cite{Semenov-2010}, do provide the first-principle
expressions for computing the thermodiffusion constant. These,
however, require the knowledge of the intermolecular potentials
and structural properties of the solvent. Hence, to obtain a
reliable estimate for the non-Fickian parameter, extensive
numerical simulations by means of Molecular Dynamics or Monte
Carlo are needed, which is beyond the scope of the present study.
Presently, however, we can prospect the value of $\alpha_T$ from a
simplified model of a large Brownian chemically-inert particle in
non-isothermal liquids. With this model, $\alpha_T$ is controlled
merely by the volume per one guest molecule in the solution and
the guest molecule mass; employing Buckingham's $\pi$-theorem of
dimensional analysis~\cite{Buckingham-1914} one can make an
interpolation of the data for dilute aqueous solutions of paraffin
alcohols: $\alpha_T\approx1.5$ for
methanol~\cite{Tichacek-Kmak-Drickamer-1956}, $\alpha_T\approx3.0$
for ethanol~\cite{Kita-Wiegand-Luettmer-Strathmann-2004}, and
$\alpha_T\approx4.5$ for
isopropanol~\cite{Poty-Legros-Thomaes-1974}. This interpolation
yields $\alpha_T\approx1.8$ for methane. Thus,
$\beta^\mathrm{CH_4}\approx-2.5$ for aqueous solutions in the
presence of geothermal gradient $G=40\,{\rm K/km}$, and
$\beta^\mathrm{CH_4}\approx-2.3$ for $G=60\,{\rm K/km}$.

For the case of carbon dioxide the non-Fickian drift affects the
solute flux much more weakly than for methane (Fig.~\ref{fig4}).
Under shallow water bodies, ${\rm CO_2}$ diffuses upwards in upper
layers of sediments, and downwards for deep layers.

\section{Conclusion}\label{sec:Concl}
We have developed the theoretical description of the process of
diffusive migration of a dissolved gas in bubbly liquids, where
bubbles are immovably trapped by a porous matrix, as in a seabed
or terrestrial aquifers. The effect of temperature inhomogeneity
across the system and the gravity force are accounted for. The
theory is employed for treatment of typical geological systems
with the hydrostatic pressure distribution and the geothermal
gradient.

For non-high pressure the diffusion flux and bubble mass evolution
are governed by Eqs.~(\ref{eq-DBL-02}) and (\ref{eq-DBL-04}),
respectively; Fig.~\ref{fig2} presents the diagram of diffusive
regimes derived for these equations. For high pressure the system
is comprehensively described by Eqs.~(\ref{eq-DBL-01}),
(\ref{eq-ScP-05}), and (\ref{eq-ScP-06}) with parameters provided
in Tab.~\ref{params} for four typical gases: nitrogen, oxygen,
carbon dioxide, and methane. The high-pressure solubility
(\ref{eq-ScP-05}) has been derived from the standard scaled
particle theory~\cite{Pierotti-1976} with adoption of van der
Waals equation of state.

The ``non-Fickian'' corrections---thermodiffusion and
gravitational segregation---appear to play an essential role in
migration of methane in the seabed, even being able to create the
gaseous methane accumulation zone in sediments (Fig.~\ref{fig3}).
For carbon dioxide the non-Fickian effects are not as significant:
they do not cause the qualitative change of behavior but can only
shift the boundary of the $\mathrm{CO_2}$-capture zone by not more
than $75\,{\rm m}$, which is $15\%$ of the characteristic depth
(Fig.~\ref{fig4}). Unfortunately, the precise values of the
thermodiffusion constant of aqueous solutions of methane or carbon
dioxide are not found in the literature, and one can only estimate
their values as we do in this paper. The practical significance of
the thermodiffusion effect, which has become apparent for the
evolution of geological systems, necessitates the experimental
determination of the thermodiffusion constant for aqueous
solutions of methane and carbon dioxide.

\begin{acknowledgements}
The authors are grateful to A.~N.\ Gorban, D.~V.\ Lyubimov, C.~A.\
Rochelle, J.~Levesley, J.~Rees, and P.~Jackson for fruitful
discussions and comments. The work has been supported by NERC
Grant no.\ NE/F021941/1.
\end{acknowledgements}

\appendix
\section{Calculation of $N_1$ and $\tilde{M}$ from experimental data}
\label{AppA}
The ratio of the volume occupied by one mole of solute molecules
in the solvent, $v_\mathrm{liq}^\mathrm{g}$, and the molar volume
of this solvent, $v_\mathrm{liq}$, [which is
$N_1=(v_\mathrm{liq}^\mathrm{g}/v_\mathrm{liq})$ in
Sec.~\ref{sec:DBL}] can be precisely derived for
$\mathrm{CO_2}$--$\mathrm{H_2O}$ and
$\mathrm{CH_4}$--$\mathrm{H_2O}$ systems from the dependence of
the solution density on its concentration, which is available in
the
literature~\cite{Hnedkovsky-Wood-Majer-1996,Garcia-2001,Duan-Hu-Li-Mao-2008}.
The molar volume of solution is
$v_\mathrm{sol}=v_\mathrm{liq}(1-X)+v_\mathrm{liq}^\mathrm{g}X$,
and the density is
\begin{eqnarray}
&&
\rho_\mathrm{sol}=\frac{M^\mathrm{host}(1-X)+M^\mathrm{g}X}{v_\mathrm{sol}}
\nonumber\\
&&
\quad =\rho_\mathrm{sol}(X=0)\left[1+\left(\frac{M^\mathrm{g}}{M^\mathrm{host}}
 -\frac{v_\mathrm{liq}^\mathrm{g}}{v_\mathrm{liq}}\right)X+O(X^2)\right],
\nonumber
\end{eqnarray}
where $M^\mathrm{host}$ and $M^\mathrm{g}$ are the molar masses of
the solvent and solute molecules, respectively;
$\rho_\mathrm{sol}(X=0)=\rho_\mathrm{liq}$ is the density of a
pure liquid (solvent). Hence,
\begin{equation}
\frac{v_\mathrm{liq}^\mathrm{g}}{v_\mathrm{liq}}
 =\frac{M^\mathrm{g}}{M^\mathrm{host}}
 -\frac{1}{\rho_\mathrm{liq}}\frac{\partial\rho_\mathrm{sol}}{\partial X},
\quad
 \tilde{M}=\frac{M^\mathrm{host}}{\rho_\mathrm{liq}}
 \frac{\partial\rho_\mathrm{sol}}{\partial X}.
\label{eq-v2v1-01}
\end{equation}
The results of the implementation of the latter relation to the
experimental data
from~\cite{Hnedkovsky-Wood-Majer-1996,Garcia-2001} are summarized
in the three last columns of Table~\ref{params}.

\section{Aqueous solubility of certain gases}
\label{AppB}
Here we wish to present some detail of our calculations, outlined
in previous sections. To evaluate the solubility at high pressure
we make a straightforward generalization of the standard scaled
particle theory. Namely, instead of the ideal-gas equation of
state we use the van der Waals equation for the vapor phase of the
dissolved gas. We find the chemical potential of the solute  and
calculate solubility for the gases, most important for
applications---oxygen, nitrogen, carbon dioxide, and methane.

\subsection{Chemical potential in the vapor phase}\label{sec:ChemPot}
To introduce notations and recall basic relations we start from an
ideal gas. The Helmholtz free energy, $F(T,V,N)$, the Gibbs free
energy, $G(T,P,N)=F+PV=F-V(\partial F/\partial V)$, and the
chemical potential, $\mu_\mathrm{vap}^\mathrm{g}(T,P) =
\partial G (T,P,N) / \partial N$, read for an ideal gas,
\begin{eqnarray}
F&=&-Nk_\mathrm{B}T\ln \frac{eV}{N}  +Nk_\mathrm{B}T \ln
\frac{\Lambda^3}{Z(T)}\,,
 \label{eq-ChP-01} \\
G&=&Nk_\mathrm{B}T\ln\frac{P}{k_\mathrm{B}T} +Nk_\mathrm{B}T \ln
\frac{\Lambda^3}{Z(T)}\,,
 \label{eq-ChP-02} \\
\mu_\mathrm{vap}^\mathrm{g}&=& k_\mathrm{B}T\ln \left[\frac{P}{k_\mathrm{B}T}\,
  \frac{\Lambda^3}{Z(T)} \right]\,,
 \label{eq-ChP-03}
\end{eqnarray}
where $N$ is the number  of molecules, $V$ is the  volume, $P$ is
pressure,  $k_\mathrm{B}$ is the Boltzmann constant, $\Lambda=
\sqrt{2 \pi
\hbar^2 /mk_\mathrm{B}T }$ is the thermal de Broglie wavelength of a gas
molecule of mass $m$, $\hbar$ is the Planck constant divided by
$2\pi$ and $Z(T)$ is the partition function of the internal
degrees of freedom of the molecule. It depends on the molecular
structure and temperature.

Now, we address  the chemical potential of  real gases, taking
into account the finite volume fraction of the gas molecules and
their attractive interactions. The simplest way to do this is to
use the van der Waals model. We assume that the heat capacity
$c_V$ does not depend on temperature. This is justified for the
range of temperatures of interest for many gases, including
methane. Then the Helmholtz and Gibbs free energies read,
e.g.~\cite{Landau-Lifshitz-V},
\begin{eqnarray}
F(T,V,N)&=&F_\mathrm{id.}-Nk_\mathrm{B}T\ln\left(1-nb \right)-N n
a, \label{eq-VdW-01}
\\[5pt]
G(T,P,N)&=& G_\mathrm{id.}-Nk_\mathrm{B}T\ln\left(1-nb\right)
\nonumber\\
&&\qquad
 {}-2Nna+Nk_\mathrm{B}T\frac{nb}{1-nb} \, .
\label{eq-VdW-03}
\end{eqnarray}
Here $F_\mathrm{id.}$  and $G_\mathrm{id.}$ are the respective
ideal parts and $a$ and $b$ are the van der Waals constants (see
Tab.~\ref{params} for the specific values). Note that above the
critical point of a gas ($T_c=-82.7\,^{\circ}\mathrm{C}$ for
methane), $n=n(P)$ is a univocal function of pressure, which may
be obtained from the van der Waals equation of state:
\begin{equation}
P =\frac{nk_\mathrm{B}T}{1-nb}-an^2\,.
\label{eq-VdW-02}
\end{equation}
Finally, the chemical potential of the  van der Waals gas reads,
\begin{eqnarray}
\mu_\mathrm{vap}^\mathrm{g}(T,P)= k_\mathrm{B}T \left[ \ln\left( \frac{n \Lambda^3}{Z(T)} \right)
 -\ln(1-bn)
\right.\quad\nonumber\\
\left. {}-\frac{2an}{k_\mathrm{B}T}+\frac{nb}{1-nb} \right].
\label{eq-VdW-04}
\end{eqnarray}
This result can be found in~\cite{Hirschfelder-Curtiss-Bird-1954}.

%%%%%%%%%%%%%%%%%%%%%%%%%%%%%%%%%%%%%%%%%%%%%%%%%%%%%%%%%%%%%%%%%%%%%%%
\begin{table*}[!t]
\caption{Parameters of gases and their aqueous solutions.
Values of $\sigma$, $B$, and $G_i$ are derived from the
theoretical fitting of the experimental data presented in
Fig.~\ref{fig5}. Parameters in the three last columns are derived
from experimental
data~\cite{Hnedkovsky-Wood-Majer-1996,Garcia-2001} with
Eq.~(\ref{eq-v2v1-01}).}
\begin{center}
\begin{tabular}{|c|c|c|c|c|c|c|c|c|c|}
\hline
\qquad&
 $\displaystyle a,\ \displaystyle {\rm\frac{m^6Pa}{mol^2}}$ &
 $\displaystyle b,\ \displaystyle 10^{-5}{\rm\frac{m^3}{mol}}$ &
 $\sigma$, ${\rm \AA}$ &
 $A(y)$ &
 $\displaystyle B,\ \displaystyle 10^{-6}{\rm\frac{K}{Pa}}$ &
 $\displaystyle \frac{G_i}{k_\mathrm{B}}$, ${\rm K}$ &
 $\displaystyle\frac{1}{\rho_0}\frac{\partial\rho_\mathrm{sol}}{\partial X}$ &
 $N_1\equiv\displaystyle\frac{v_\mathrm{liq}^\mathrm{g}}{v_\mathrm{liq}}$ &
 $ \displaystyle\tilde{M},\ \displaystyle {\rm \frac{g}{mol}}$ \\
\hline
$\mathrm{O_2}$  & $0.1378$ & $3.183$ & $3.03$ & $6.38$ & $1.056$ & $-873$ & --- & --- & --- \\
\hline
$\mathrm{N_2}$  & $0.1408$ & $3.913$ & $3.15$ & $6.77$ & $1.182$ & $-781$ & --- & --- & --- \\
\hline
$\mathrm{CO_2}$ & $0.3640$ & $4.267$ & $3.02$ & $6.36$ & $1.050$ & $-1850$ &  $0.54$ & $1.91$ & $9.7$ \\
\hline
$\mathrm{CH_4}$ & $0.2283$ & $4.278$ & $3.27$ & $7.18$ & $1.321$ & $-1138$ & $-1.35$ & $2.24$ & $-24.3$ \\
\hline
$\mathrm{H_2O}$ & --- & --- & $2.77$ & --- & --- & --- & --- & --- & --- \\
\hline
\end{tabular}
\end{center}
\label{params}
\end{table*}
%%%%%%%%%%%%%%%%%%%%%%%%%%%%%%%%%%%%%%%%%%%%%%%%%%%%%%%%%%%%%%%%%%%%%%%

%%%%%%%%%%%%%%%%%%%%%%%%%%%%%%%%%%%%%%%%%%%%%%%%%%%%%%%%%%%%%%%
\begin{figure*}[t]
\center{
 {\sf (a)\hspace{-3mm}}
 \includegraphics[width=0.323\textwidth]%
 {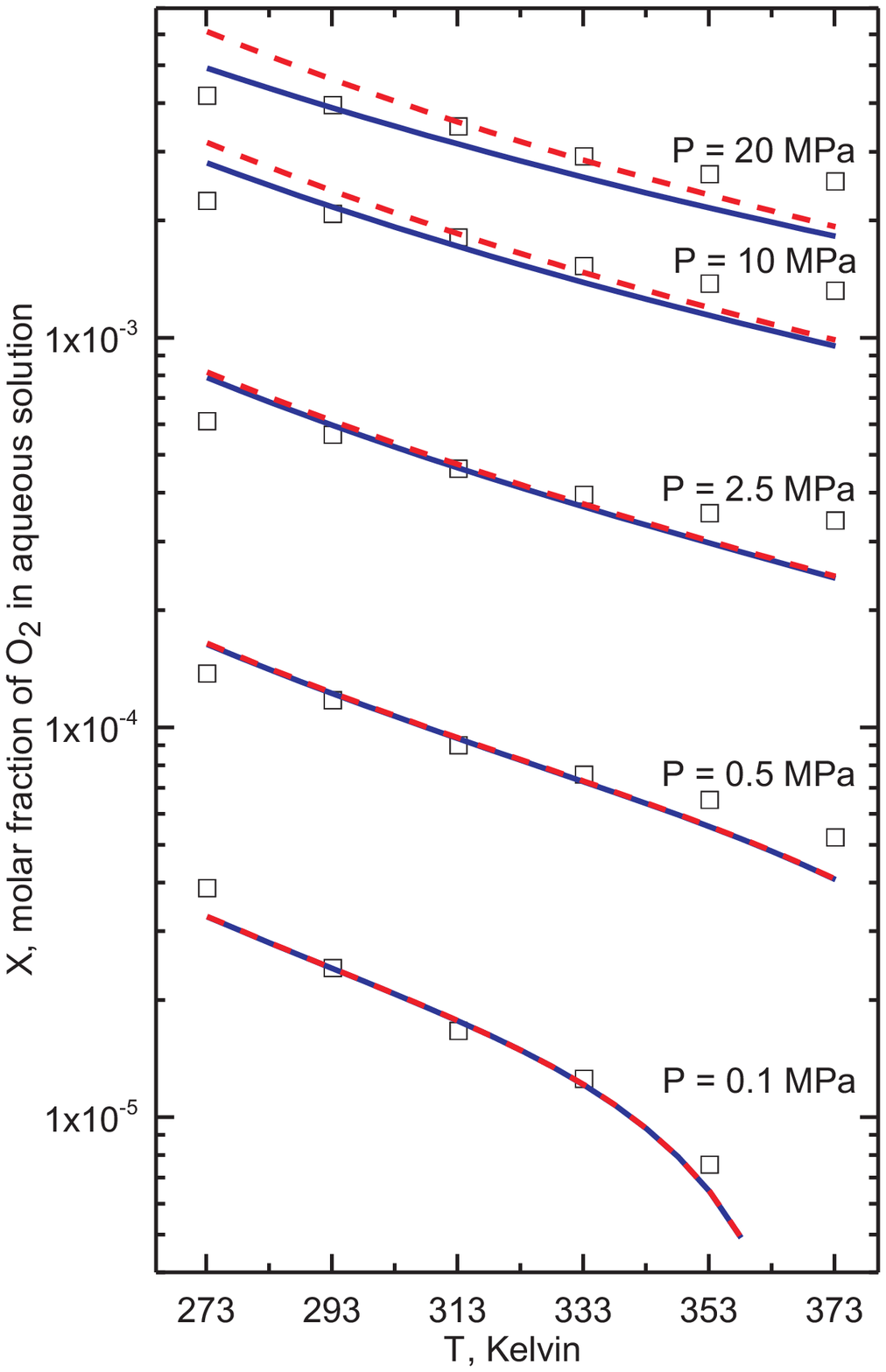}
 \qquad\qquad\qquad
 {\sf (b)\hspace{-3mm}}
 \includegraphics[width=0.323\textwidth]%
 {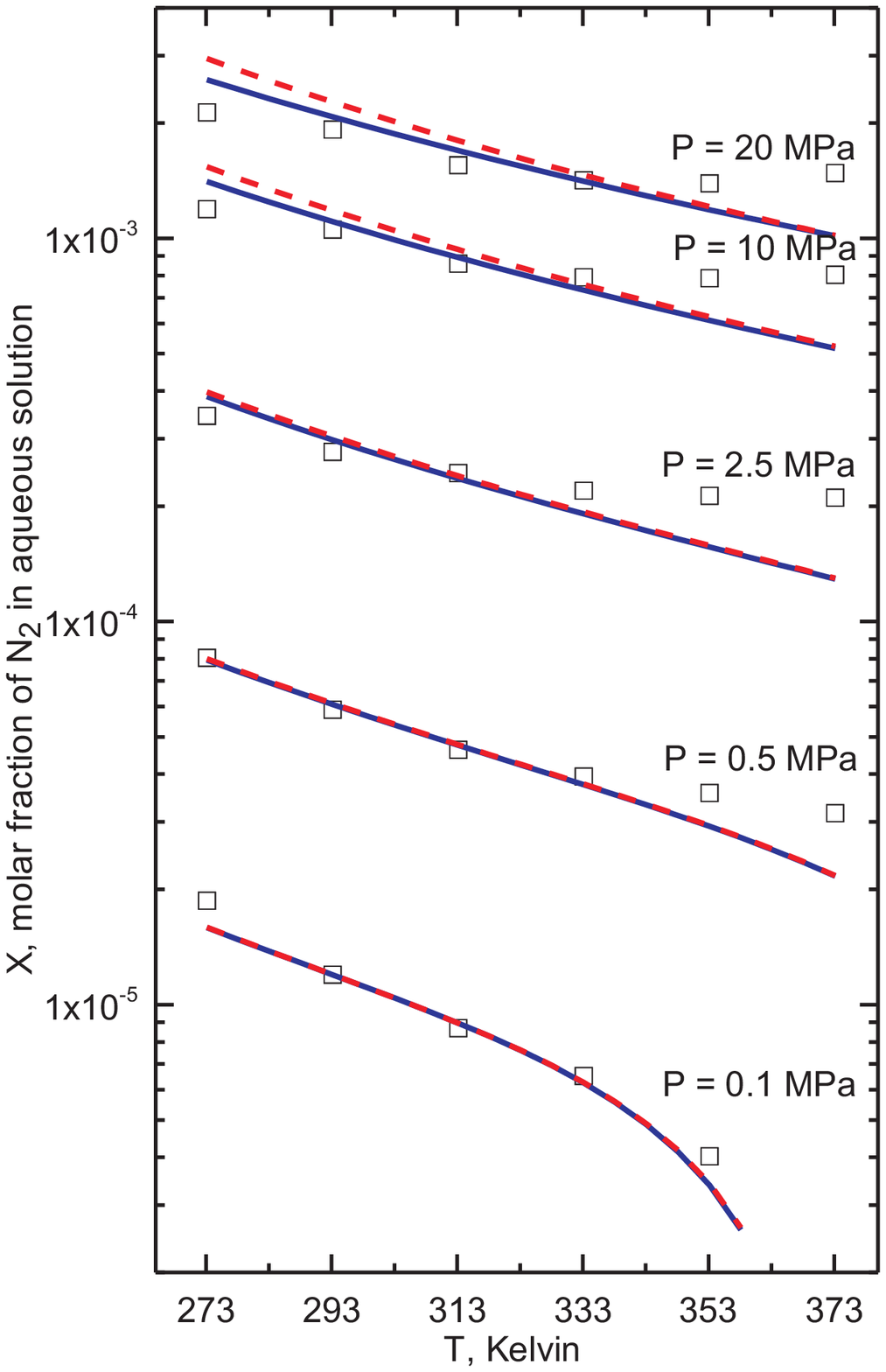}
}

\vspace{5pt}
\center{
 {\sf (c)\hspace{-3mm}}
 \includegraphics[width=0.323\textwidth]%
 {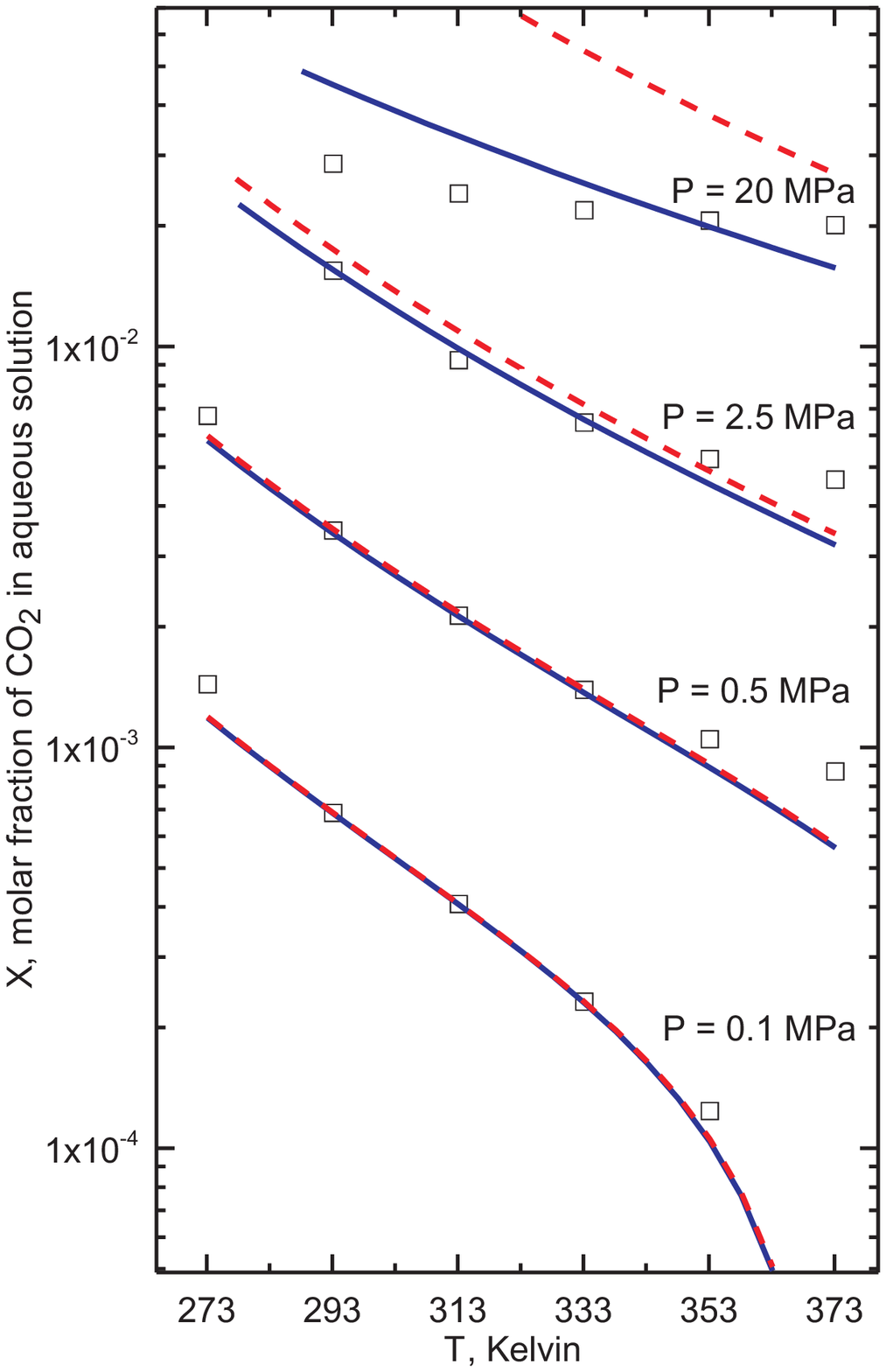}
 \qquad\qquad\qquad
 {\sf (d)\hspace{-3mm}}
 \includegraphics[width=0.323\textwidth]%
 {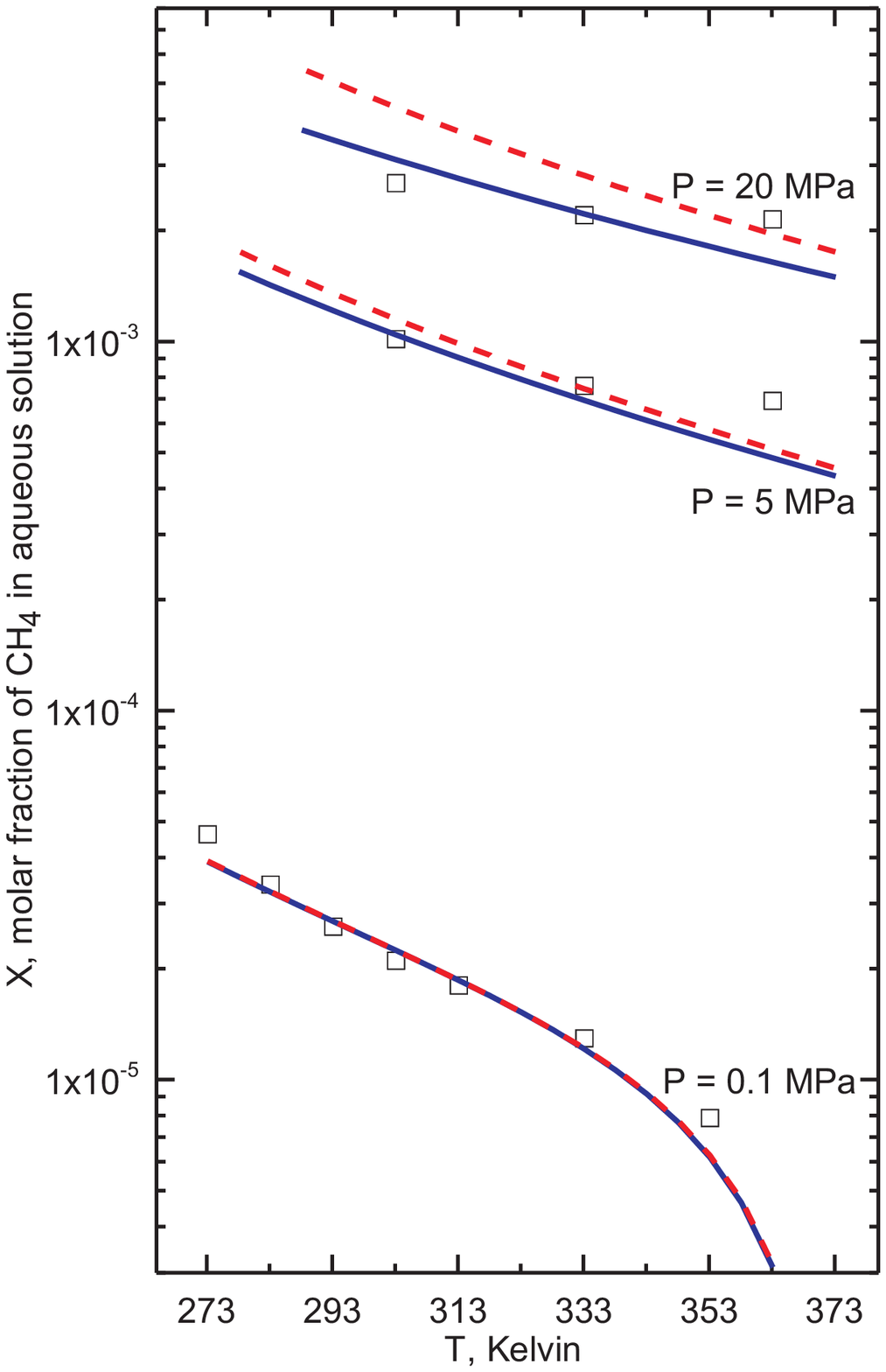}
}
  \caption{(Color online) Solubility
of $\mathrm{O_2}$ (a), $\mathrm{N_2}$ (b), $\mathrm{CO_2}$ (c),
and $\mathrm{CH_4}$ (d) in water at specified pressure. Squares
represent experimental data
(\cite{Baranenko-1990a,Baranenko-1990b} for nitrogen, oxygen, and
carbon dioxide and~\cite{Yamomoto-et_al-1976,Duan-Mao-2006} for
methane); red dashed lines: results of the scaled particle theory
for vapor phase assumed to be an ideal gas, Eq.~(\ref{eq-ScP-04});
blue solid lines: the scaled particle theory with van der Waals'
equation of state for the vapor phase, Eq.~(\ref{eq-ScP-05}).
 }
  \label{fig5}
\end{figure*}
%%%%%%%%%%%%%%%%%%%%%%%%%%%%%%%%%%%%%%%%%%%%%%%%%%%%%%%%%%%%%%%

\subsection{Solubility of gas: Scaled particle theory}\label{sec:ScalPart}
To evaluate the solubility of a gas in a solvent (water) we employ
scaled particle theory (see e.g. the review~\cite{Pierotti-1976}
for detail). When the concentration of the dissolved gas reaches
solubility, the system is in thermodynamic equilibrium, so that
the chemical potential of the gas in the vapor phase
$\mu_\mathrm{vap}^\mathrm{g}$ is equal to this in the solution
$\mu_\mathrm{liq}^\mathrm{g}$. According to the scaled particle
theory~\cite{Pierotti-1976},
\begin{equation}
\mu_\mathrm{liq}^\mathrm{g}=G_c+G_i+k_\mathrm{B}T\ln(\Lambda^3/Z)
 +k_\mathrm{B}T\ln(XN_\mathrm{A}/v_\mathrm{liq}),
\label{eq-ScP-01}
\end{equation}
where $\Lambda$ and $Z$ are the de Broglie wavelength and the
partition functions per molecule for the internal degrees of
freedom for the solute, $X$ is the molar fraction of the gas,
$G_c$ is the work of creation of a cavity for a guest molecule in
the solvent, $G_i$ is the interaction energy between a solute
molecule and the surrounding solvent molecules, $N_\mathrm{A}$ is
the Avogadro number, and $v_\mathrm{liq}$ is the molar volume of
the solvent particles (e.g.
$v_{\mathrm{H}_2\mathrm{O}}=17.95\,{\rm cm^3/mol}$ at atmospheric
pressure and $T=300\,{\rm K}$). The cavity formation work
reads~\cite{Pierotti-1976}
\begin{eqnarray}
&&
\frac{G_c}{k_\mathrm{B}T}=-\ln(1-y)+\frac{3y}{1-y}
\frac{\sigma_\mathrm{g}}{\sigma_\mathrm{liq}}
\nonumber\\
&&
 \qquad{}+\left[\frac{3y}{1-y}+\frac{9}{2}\left(\frac{y}{1-y}\right)^2\right]
 \left(\frac{\sigma_\mathrm{g}}{\sigma_\mathrm{liq}}\right)^2
\nonumber\\
&&
 \qquad{}+\frac{yP}{n_\mathrm{liq}k_\mathrm{B}T}
 \left(\frac{\sigma_\mathrm{g}}{\sigma_\mathrm{liq}}\right)^3
 \equiv A(y)+B(y)\frac{P}{T},
\label{eq-ScP-02}
\end{eqnarray}
where $\sigma_\mathrm{liq}$ and $\sigma_\mathrm{g}$ are
respectively the effective diameter of the solvent (liquid) and
solute (gas) molecules,  $n_\mathrm{liq}$ is the number density of
solvent, and $y=(\pi/6)n_\mathrm{liq}\sigma_\mathrm{liq}^3$ (e.g.,
$y_\mathrm{H_2O}=0.371$ at $T=300\,{\rm K}$~\cite{Pierotti-1976}).
There exist numerous models for the  interaction energy $G_i$
(e.g.~\cite{Pierotti-1976,Hildebrand-Scott-1964}); for the present
study it is enough, however,  to mention that it is almost
independent of pressure and temperature.

For the vapor phase, the chemical potential is given either by
Eq.~(\ref{eq-ChP-03}) for an ideal gas or by Eq.~(\ref{eq-VdW-04})
with (\ref{eq-VdW-02}) for $n=n(P)$.
\begin{equation}
\mu_\mathrm{vap}^\mathrm{g}=k_\mathrm{B}T\ln(\Lambda^3/Z)
 +k_\mathrm{B}T\ln(f/k_\mathrm{B}T),
\label{eq-ScP-03}
\end{equation}
where $f$ is the fugacity of the gas molecules in the vapor phase.
It is equal to the pressure for a one-component gas, and to the
partial pressure for a gas mixture. In the case of interest,
$f=P(1-Y)$, where $Y$ is the molar fraction of solvent vapor in
the gaseous phase; it is determined by Eq.~(\ref{eq-ScP-06})
below. Equating Eqs.~(\ref{eq-ScP-01}) and (\ref{eq-ScP-03}) for
the chemical potential, one finds the equilibrium molar fraction
of the dissolved gas, that is the {\em solubility},  $X^{(0)}$.
For an ideal gas one obtains, using Eq.~(\ref{eq-ScP-03}) together
with the ideal gas equation of state
\begin{eqnarray}
&&\hspace{-20pt}
X^{(0)}=\frac{(1-Y)P\,v_\mathrm{liq}}{RT}
  \exp\left[-\frac{G_c+G_i}{k_\mathrm{B}T}\right]
\nonumber\\
&&=\frac{(1-Y)P\,v_\mathrm{liq}}{RT}
  \exp\left[-A(y)-B\frac{P}{T}-\frac{G_i}{k_\mathrm{B}T}\right].
\label{eq-ScP-04}
\end{eqnarray}
For a van der Waals gas, the real gas equation of state,
Eq.~(\ref{eq-VdW-02}), is to be employed. This yields
\begin{eqnarray}
X^{(0)}=\frac{(1-Y)n\,v_\mathrm{liq}}{(1-nb)N_\mathrm{A}}
 \exp\left[-\frac{G_c+G_i}{k_\mathrm{B}T}
 -\frac{2an}{k_\mathrm{B}T}+\frac{1}{1-nb}
\right]
\nonumber\\
=\frac{(1-Y)n\,v_\mathrm{liq}}{(1-nb)N_\mathrm{A}}
 \exp\left[-A(y)-B\frac{P}{T}-\frac{G_i}{k_\mathrm{B}T}
\right.\quad
\nonumber\\
\left.
 {}-\frac{2an}{k_\mathrm{B}T}+\frac{1}{1-nb}\right].\qquad
\label{eq-ScP-05}
\end{eqnarray}

When liquid with a dissolved gas is in equilibrium with its vapor
phase, the vapor, including the vapor in bubbles,  contains both
components.  Let us evaluate the solvent molar fraction $Y$ in the
vapor phase. The enthalpy variation $dH=c_PdT+VdP=T\,dS+VdP$.
Hence,
 $H=H_0+\int_{T_0}^Tc_P(T_1)\,dT_1+\int_{P_0}^PV(T,P)\,dP$,
 $S=S_0+\int_{T_0}^Tc_P(T_1)T_1^{-1}dT_1$,
and Gibbs free energy $G=H-TS
 =H_0+\int_{T_0}^Tc_P(T_1)\,dT_1+\int_{P_0}^PV(T,P)\,dP
 -T\,S_0-T\int_{T_0}^Tc_P(T_1)T_1^{-1}dT_1$.

For liquid water, $c_P(T)$ is nearly constant and
\begin{eqnarray}
G_\mathrm{liq}^\mathrm{host}=H_{\mathrm{liq},0}^\mathrm{host}
 +c_{P,\mathrm{liq}}^\mathrm{host}(T-T_0)
 +\frac{v_\mathrm{liq}}{N_A}(P-P_0)\qquad
\nonumber\\
 {}-T\,S_\mathrm{liq,0}^\mathrm{host}
 -c_{P,\mathrm{liq}}^\mathrm{host}T\ln\frac{T}{T_0}\,;
\nonumber
\end{eqnarray}
for the vapor phase,
\begin{eqnarray}
G_\mathrm{vap}^\mathrm{host}=H_{\mathrm{vap},0}^\mathrm{host}
 +c_{P,\mathrm{vap}}^\mathrm{host}(T-T_0)
 +k_\mathrm{B}T\ln\frac{YP}{P_0}\qquad
\nonumber\\
{}-T\,S_\mathrm{vap,0}^\mathrm{host}
 -c_{P,\mathrm{vap}}^\mathrm{host}T\ln\frac{T}{T_0}.
\nonumber
\end{eqnarray}
Here $Y$ appears owing to the fact that the part of entropy
related to the translational degrees of freedom is
$k_\mathrm{B}\ln(V/Y\Lambda^3)$ but not
$k_\mathrm{B}\ln(V/\Lambda^3)$ as for a pure steam.

In equilibrium, the chemical potential in the liquid and vapor
phases are equal,
$G_\mathrm{vap}^\mathrm{host}=G_\mathrm{liq}^\mathrm{host}$, and
\begin{eqnarray}
Y^\mathrm{H_2O}=
 \frac{P_0}{P}
 \left(\frac{T}{T_0}\right)^{\textstyle\frac{\Delta c_P^\mathrm{H_2O}}{k_\mathrm{B}}}
 \exp\left[\frac{v_\mathrm{liq}(P-P_0)}{RT}
 \right.\quad
\nonumber\\
 \left.
 {}-\frac{\Delta H_0^\mathrm{H_2O}-\Delta c_P^\mathrm{H_2O}T_0}{k_\mathrm{B}}
 \left(\frac{1}{T}-\frac{1}{T_0}\right)\right],
\label{eq-ScP-06}
\end{eqnarray}
where
 $\Delta c_P^\mathrm{H_2O}=c_{P,\mathrm{vap}}^\mathrm{H_2O}-c_{P,\mathrm{liq}}^\mathrm{H_2O}$,
 $c_{P,\mathrm{vap}}^\mathrm{H_2O}/k_\mathrm{B}=4.09$,
 $c_{P,\mathrm{liq}}^\mathrm{H_2O}/k_\mathrm{B}=9.09$,
the enthalpy of vaporization
 $\Delta H_0^\mathrm{H_2O}/k_\mathrm{B}=4892{\rm K}$
at $T_0=373{\rm K}$ and $P_0=1{\rm atm}$. Eq.~(\ref{eq-ScP-06})
provides the approximate value of the saturated vapor pressure
which is nearly indistinguishable from the experimental value and
the known empiric formulae (e.g.~\cite{Goff-1957}).

Table \ref{params} presents parameters $\sigma$, $B$, and $G_i$
derived from experimental
data~\cite{Baranenko-1990a,Baranenko-1990b,Yamomoto-et_al-1976,Duan-Mao-2006}
with the employment of Eqs.~(\ref{eq-ScP-04}) and
(\ref{eq-ScP-06}) for an ideal gas and Eqs.~(\ref{eq-ScP-05}) and
(\ref{eq-ScP-06}) for van der Waals' model. The agreement between
the theory and experimental data can be assessed from
Fig.~\ref{fig5}. Notably, van der Waals' model provides an
accurate description in the entire range of parameters relevant to
our consideration. One should keep in mind that van der Waals'
model provides only a qualitative description for liquid--gas
transition and, therefore, the formulae we present are not
applicable for the description of dissolution of any gas at its
liquefaction conditions.

\bibliographystyle{apsrev}

\begin{thebibliography}{37}
\expandafter\ifx\csname natexlab\endcsname\relax\def\natexlab#1{#1}\fi
\expandafter\ifx\csname bibnamefont\endcsname\relax
  \def\bibnamefont#1{#1}\fi
\expandafter\ifx\csname bibfnamefont\endcsname\relax
  \def\bibfnamefont#1{#1}\fi
\expandafter\ifx\csname citenamefont\endcsname\relax
  \def\citenamefont#1{#1}\fi
\expandafter\ifx\csname url\endcsname\relax
  \def\url#1{\texttt{#1}}\fi
\expandafter\ifx\csname urlprefix\endcsname\relax\def\urlprefix{URL }\fi
\providecommand{\bibinfo}[2]{#2}
\providecommand{\eprint}[2][]{\url{#2}}

\bibitem[{\citenamefont{Hirschfelder et~al.}(1954)\citenamefont{Hirschfelder,
  Curtiss, and Bird}}]{Hirschfelder-Curtiss-Bird-1954}
\bibinfo{author}{\bibfnamefont{J.~O.} \bibnamefont{Hirschfelder}},
  \bibinfo{author}{\bibfnamefont{C.~F.} \bibnamefont{Curtiss}},
  \bibnamefont{and} \bibinfo{author}{\bibfnamefont{R.~B.} \bibnamefont{Bird}},
  \emph{\bibinfo{title}{The Molecular Theory of Gases and Liquids}}
  (\bibinfo{publisher}{Wiley}, \bibinfo{year}{1954}).

\bibitem[{\citenamefont{Bird et~al.}(2007)\citenamefont{Bird, Stewart, and
  Lightfoot}}]{Bird-Stewart-Lightfoot-2007}
\bibinfo{author}{\bibfnamefont{R.~B.} \bibnamefont{Bird}},
  \bibinfo{author}{\bibfnamefont{W.~E.} \bibnamefont{Stewart}},
  \bibnamefont{and} \bibinfo{author}{\bibfnamefont{E.~N.}
  \bibnamefont{Lightfoot}}, \emph{\bibinfo{title}{Transport Phenomena}}
  (\bibinfo{publisher}{Wiley}, \bibinfo{year}{2007}), \bibinfo{edition}{2nd}
  ed.

\bibitem[{\citenamefont{Yurkovetsky and Brady}(1996)}]{Yurkovetsky-Brady-1996}
\bibinfo{author}{\bibfnamefont{Y.}~\bibnamefont{Yurkovetsky}} \bibnamefont{and}
  \bibinfo{author}{\bibfnamefont{J.~F.} \bibnamefont{Brady}},
  \bibinfo{journal}{Phys.\ Fluids} \textbf{\bibinfo{volume}{8}},
  \bibinfo{pages}{881} (\bibinfo{year}{1996}).

\bibitem[{\citenamefont{Archer}(2007)}]{Archer-2007}
\bibinfo{author}{\bibfnamefont{D.}~\bibnamefont{Archer}},
  \bibinfo{journal}{Biogeosciences} \textbf{\bibinfo{volume}{4}},
  \bibinfo{pages}{521} (\bibinfo{year}{2007}).

\bibitem[{\citenamefont{Davie and Buffett}(2001)}]{Davie-Buffett-2001}
\bibinfo{author}{\bibfnamefont{M.~K.} \bibnamefont{Davie}} \bibnamefont{and}
  \bibinfo{author}{\bibfnamefont{B.~A.} \bibnamefont{Buffett}},
  \bibinfo{journal}{J.\ Geophys.\ Res.} \textbf{\bibinfo{volume}{106}},
  \bibinfo{pages}{497} (\bibinfo{year}{2001}).

\bibitem[{\citenamefont{Davie and Buffett}(2003)}]{Davie-Buffett-2003}
\bibinfo{author}{\bibfnamefont{M.~K.} \bibnamefont{Davie}} \bibnamefont{and}
  \bibinfo{author}{\bibfnamefont{B.~A.} \bibnamefont{Buffett}},
  \bibinfo{journal}{J.\ Geophys.\ Res.} \textbf{\bibinfo{volume}{108}},
  \bibinfo{pages}{2495} (\bibinfo{year}{2003}).

\bibitem[{\citenamefont{Garg et~al.}(2008)\citenamefont{Garg, Pritchett, Katoh,
  Baba, and Fujii}}]{Garg_etal-2008}
\bibinfo{author}{\bibfnamefont{S.~K.} \bibnamefont{Garg}},
  \bibinfo{author}{\bibfnamefont{J.~W.} \bibnamefont{Pritchett}},
  \bibinfo{author}{\bibfnamefont{A.}~\bibnamefont{Katoh}},
  \bibinfo{author}{\bibfnamefont{K.}~\bibnamefont{Baba}}, \bibnamefont{and}
  \bibinfo{author}{\bibfnamefont{T.}~\bibnamefont{Fujii}},
  \bibinfo{journal}{J.\ Geophys.\ Res.} \textbf{\bibinfo{volume}{113}},
  \bibinfo{pages}{B01201} (\bibinfo{year}{2008}).

\bibitem[{\citenamefont{Haacke et~al.}({2008})\citenamefont{Haacke, Westbrook,
  and Riley}}]{Haacke-Westbrook-Riley-2008}
\bibinfo{author}{\bibfnamefont{R.~R.} \bibnamefont{Haacke}},
  \bibinfo{author}{\bibfnamefont{G.~K.} \bibnamefont{Westbrook}},
  \bibnamefont{and} \bibinfo{author}{\bibfnamefont{M.~S.} \bibnamefont{Riley}},
  \bibinfo{journal}{{J.\ Geophys.\ Res.}} \textbf{\bibinfo{volume}{{113}}}
  \bibinfo{pages}{B05104}(\bibinfo{year}{{2008}}).

\bibitem[{\citenamefont{Donaldson et~al.}(1997)\citenamefont{Donaldson, Istok,
  Humphrey, O'Reilly, Hawelka, and Mohr}}]{Donaldson-etal-1997}
\bibinfo{author}{\bibfnamefont{J.~H.} \bibnamefont{Donaldson}},
  \bibinfo{author}{\bibfnamefont{J.~D.} \bibnamefont{Istok}},
  \bibinfo{author}{\bibfnamefont{M.~D.} \bibnamefont{Humphrey}},
  \bibinfo{author}{\bibfnamefont{K.~T.} \bibnamefont{O'Reilly}},
  \bibinfo{author}{\bibfnamefont{C.~A.} \bibnamefont{Hawelka}},
  \bibnamefont{and} \bibinfo{author}{\bibfnamefont{D.~H.} \bibnamefont{Mohr}},
  \bibinfo{journal}{Ground Water} \textbf{\bibinfo{volume}{35}},
  \bibinfo{pages}{270} (\bibinfo{year}{1997}).

\bibitem[{\citenamefont{Donaldson et~al.}(1998)\citenamefont{Donaldson, Istok,
  and O'Reilly}}]{Donaldson-etal-1998}
\bibinfo{author}{\bibfnamefont{J.~H.} \bibnamefont{Donaldson}},
  \bibinfo{author}{\bibfnamefont{J.~D.} \bibnamefont{Istok}}, \bibnamefont{and}
  \bibinfo{author}{\bibnamefont{O'Reilly}}, \bibinfo{journal}{Ground Water}
  \textbf{\bibinfo{volume}{36}}, \bibinfo{pages}{133} (\bibinfo{year}{1998}).

\bibitem[{\citenamefont{Holloway}(1997)}]{Holloway-1997}
\bibinfo{author}{\bibfnamefont{S.}~\bibnamefont{Holloway}},
  \bibinfo{journal}{Energy Conversion and Management}
  \textbf{\bibinfo{volume}{38}}, \bibinfo{pages}{S193 } (\bibinfo{year}{1997}),
  \bibinfo{note}{proceedings of the Third
  International Conference on Carbon Dioxide Removal}.

\bibitem[{\citenamefont{Rochelle et~al.}(2009)\citenamefont{Rochelle, Camps,
  Long, Milodowski, Bateman, Gunn, Jackson, Lovell, and
  Rees}}]{Rochelle_etal-2009}
\bibinfo{author}{\bibfnamefont{C.~A.} \bibnamefont{Rochelle}},
  \bibinfo{author}{\bibfnamefont{A.~P.} \bibnamefont{Camps}},
  \bibinfo{author}{\bibfnamefont{D.}~\bibnamefont{Long}},
  \bibinfo{author}{\bibfnamefont{A.}~\bibnamefont{Milodowski}},
  \bibinfo{author}{\bibfnamefont{K.}~\bibnamefont{Bateman}},
  \bibinfo{author}{\bibfnamefont{D.}~\bibnamefont{Gunn}},
  \bibinfo{author}{\bibfnamefont{P.}~\bibnamefont{Jackson}},
  \bibinfo{author}{\bibfnamefont{M.~A.} \bibnamefont{Lovell}},
  \bibnamefont{and} \bibinfo{author}{\bibfnamefont{J.}~\bibnamefont{Rees}},
  \bibinfo{journal}{Geological Society, London, Special Publications}
  \textbf{\bibinfo{volume}{319}}, \bibinfo{pages}{171} (\bibinfo{year}{2009}).

\bibitem[{\citenamefont{{Lyubimov} et~al.}(2009)\citenamefont{{Lyubimov},
  {Shklyaev}, {Lyubimova}, and {Zikanov}}}]{Lyubimov-etal-2009}
\bibinfo{author}{\bibfnamefont{D.~V.} \bibnamefont{{Lyubimov}}},
  \bibinfo{author}{\bibfnamefont{S.}~\bibnamefont{{Shklyaev}}},
  \bibinfo{author}{\bibfnamefont{T.~P.} \bibnamefont{{Lyubimova}}},
  \bibnamefont{and}
  \bibinfo{author}{\bibfnamefont{O.}~\bibnamefont{{Zikanov}}},
  \bibinfo{journal}{Phys.\ Fluids} \textbf{\bibinfo{volume}{21}},
  \bibinfo{pages}{014105} (\bibinfo{year}{2009}).

\bibitem[{\citenamefont{Sor\'et}(1879)}]{Soret-1879}
\bibinfo{author}{\bibfnamefont{C.}~\bibnamefont{Sor\'et}},
  \bibinfo{journal}{Archives des Sciences Physiques et Naturelles de Gen\`eve}
  \textbf{\bibinfo{volume}{2}}, \bibinfo{pages}{48} (\bibinfo{year}{1879}).

\bibitem[{\citenamefont{Jones and Furry}(1946)}]{Jones-Furry-1946}
\bibinfo{author}{\bibfnamefont{R.~C.} \bibnamefont{Jones}} \bibnamefont{and}
  \bibinfo{author}{\bibfnamefont{W.~H.} \bibnamefont{Furry}},
  \bibinfo{journal}{Rev.\ Mod.\ Phys.} \textbf{\bibinfo{volume}{18}},
  \bibinfo{pages}{151} (\bibinfo{year}{1946}).

\bibitem[{\citenamefont{Richter}(1972)}]{Richter-1972}
\bibinfo{author}{\bibfnamefont{J.}~\bibnamefont{Richter}},
  \bibinfo{journal}{Geoderma} \textbf{\bibinfo{volume}{8}}, \bibinfo{pages}{95}
  (\bibinfo{year}{1972}).


\bibitem[{\citenamefont{Saffman}(1959)}]{Saffman-1959}
\bibinfo{author}{\bibfnamefont{P.~G.} \bibnamefont{Saffman}},
  \bibinfo{journal}{J.\ Fluid Mech.} \textbf{\bibinfo{volume}{6}},
  \bibinfo{pages}{321} (\bibinfo{year}{1959}).

\bibitem[{\citenamefont{Sahimi}(1993)}]{Sahimi-1993}
\bibinfo{author}{\bibfnamefont{M.}~\bibnamefont{Sahimi}},
  \bibinfo{journal}{Rev.\ Mod.\ Phys.} \textbf{\bibinfo{volume}{65}},
  \bibinfo{pages}{1393} (\bibinfo{year}{1993}).

\bibitem[{\citenamefont{Pierotti}(1976)}]{Pierotti-1976}
\bibinfo{author}{\bibfnamefont{R.~A.} \bibnamefont{Pierotti}},
  \bibinfo{journal}{Chem.\ Rev.} \textbf{\bibinfo{volume}{76}},
  \bibinfo{pages}{717} (\bibinfo{year}{1976}).

\bibitem[{\citenamefont{Dominguez et~al.}(2000)\citenamefont{Dominguez, Bories,
  and Prat}}]{Dominguez-Bories-Prat-2000}
\bibinfo{author}{\bibfnamefont{A.}~\bibnamefont{Dominguez}},
  \bibinfo{author}{\bibfnamefont{S.}~\bibnamefont{Bories}}, \bibnamefont{and}
  \bibinfo{author}{\bibfnamefont{M.}~\bibnamefont{Prat}},
  \bibinfo{journal}{Int.\ J.\ Multiphase Flow}
  \textbf{\bibinfo{volume}{26}}, \bibinfo{pages}{1951 } (\bibinfo{year}{2000}).

\bibitem[{\citenamefont{Goldobin}(2011)}]{Goldobin-EPL-2011}
\bibinfo{author}{\bibfnamefont{D.~S.} \bibnamefont{Goldobin}},
  \bibinfo{journal}{Europhys.\ Lett.} \textbf{\bibinfo{volume}{95}},
  \bibinfo{pages}{64004} (\bibinfo{year}{2011}).


\bibitem[{\citenamefont{Semenov}(2010)}]{Semenov-2010}
\bibinfo{author}{\bibfnamefont{S.~N.} \bibnamefont{Semenov}},
  \bibinfo{journal}{Europhys.\ Lett.} \textbf{\bibinfo{volume}{90}},
  \bibinfo{pages}{56002} (\bibinfo{year}{2010}).

\bibitem[{\citenamefont{Buckingham}(1914)}]{Buckingham-1914}
\bibinfo{author}{\bibfnamefont{E.} \bibnamefont{Buckingham}},
  \bibinfo{journal}{Phys.\ Rev.} \textbf{\bibinfo{volume}{4}},
  \bibinfo{pages}{345} (\bibinfo{year}{1914}).
%On physically similar systems; illustrations of the use of
%dimensional equations

\bibitem[{\citenamefont{Tichacek et~al.}(1956)\citenamefont{Tichacek, Kmak, and
  Drickamer}}]{Tichacek-Kmak-Drickamer-1956}
\bibinfo{author}{\bibfnamefont{L.~G.}~\bibnamefont{Tichacek}},
  \bibinfo{author}{\bibfnamefont{W.~S.}~\bibnamefont{Kmak}}, \bibnamefont{and}
  \bibinfo{author}{\bibfnamefont{H.~G.}~\bibnamefont{Drickamer}},
  \bibinfo{journal}{J.\ Phys.\ Chem.}
  \textbf{\bibinfo{volume}{60}}, \bibinfo{pages}{660} (\bibinfo{year}{1956}).

\bibitem[{\citenamefont{Kita et~al.}(2004)\citenamefont{Kita, Wiegand, and
  Luettmer-Strathmann}}]{Kita-Wiegand-Luettmer-Strathmann-2004}
\bibinfo{author}{\bibfnamefont{R.}~\bibnamefont{Kita}},
  \bibinfo{author}{\bibfnamefont{S.}~\bibnamefont{Wiegand}}, \bibnamefont{and}
  \bibinfo{author}{\bibfnamefont{J.}~\bibnamefont{Luettmer-Strathmann}},
  \bibinfo{journal}{J.\ Chem.\ Phys.}
  \textbf{\bibinfo{volume}{121}}, \bibinfo{pages}{3874} (\bibinfo{year}{2004}).

\bibitem[{\citenamefont{Poty et~al.}(1974)\citenamefont{Poty, Legros, and
  Thomaes}}]{Poty-Legros-Thomaes-1974}
\bibinfo{author}{\bibfnamefont{P.}~\bibnamefont{Poty}},
  \bibinfo{author}{\bibfnamefont{J.~C.}~\bibnamefont{Legros}}, \bibnamefont{and}
  \bibinfo{author}{\bibfnamefont{G.}~\bibnamefont{Thomaes}},
  \bibinfo{journal}{Z.\ Naturforsch.\ A} \textbf{\bibinfo{volume}{29A}},
  \bibinfo{pages}{1915} (\bibinfo{year}{1974}).


\bibitem[{\citenamefont{Hnedkovsky et~al.}(1996)\citenamefont{Hnedkovsky, Wood,
  and Majer}}]{Hnedkovsky-Wood-Majer-1996}
\bibinfo{author}{\bibfnamefont{L.}~\bibnamefont{Hnedkovsky}},
  \bibinfo{author}{\bibfnamefont{R.~H.} \bibnamefont{Wood}}, \bibnamefont{and}
  \bibinfo{author}{\bibfnamefont{V.}~\bibnamefont{Majer}},
  \bibinfo{journal}{J.\ Chem.\ Thermodyn.}
  \textbf{\bibinfo{volume}{28}}, \bibinfo{pages}{125} (\bibinfo{year}{1996}).

\bibitem[{\citenamefont{Garcia}(2001)}]{Garcia-2001}
\bibinfo{author}{\bibfnamefont{J.~E.} \bibnamefont{Garcia}},
  \bibinfo{journal}{Lawrence Berkeley National Laboratory Paper} pp.
  \bibinfo{pages}{LBNL--49023} (\bibinfo{year}{2001}).

\bibitem[{\citenamefont{Duan et~al.}(2008)\citenamefont{Duan, Hu, Li, and
  Mao}}]{Duan-Hu-Li-Mao-2008}
\bibinfo{author}{\bibfnamefont{Z.}~\bibnamefont{Duan}},
  \bibinfo{author}{\bibfnamefont{J.}~\bibnamefont{Hu}},
  \bibinfo{author}{\bibfnamefont{D.}~\bibnamefont{Li}}, \bibnamefont{and}
  \bibinfo{author}{\bibfnamefont{S.}~\bibnamefont{Mao}},
  \bibinfo{journal}{{Energy \& Fuels}} \textbf{\bibinfo{volume}{22}},
  \bibinfo{pages}{1666} (\bibinfo{year}{2008}).

\bibitem[{\citenamefont{Landau and Lifshitz}(1996)}]{Landau-Lifshitz-V}
\bibinfo{author}{\bibfnamefont{L.~D.} \bibnamefont{Landau}} \bibnamefont{and}
  \bibinfo{author}{\bibfnamefont{E.~M.} \bibnamefont{Lifshitz}},
  \emph{\bibinfo{title}{Statistical Physics: Volume 5 (Course of Theoretical
  Physics)}} (\bibinfo{publisher}{A Butterworth-Heinemann Title},
  \bibinfo{year}{1996}), \bibinfo{edition}{3rd} ed.

\bibitem[{\citenamefont{Baranenko
  et~al.}(1990{\natexlab{a}})\citenamefont{Baranenko, Sysoev, Fal'kovskii,
  Kirov, Piontkovskii, and Musienko}}]{Baranenko-1990a}
\bibinfo{author}{\bibfnamefont{V.~I.} \bibnamefont{Baranenko}},
  \bibinfo{author}{\bibfnamefont{V.~S.} \bibnamefont{Sysoev}},
  \bibinfo{author}{\bibfnamefont{L.~N.} \bibnamefont{Fal'kovskii}},
  \bibinfo{author}{\bibfnamefont{V.~S.} \bibnamefont{Kirov}},
  \bibinfo{author}{\bibfnamefont{A.~I.} \bibnamefont{Piontkovskii}},
  \bibnamefont{and} \bibinfo{author}{\bibfnamefont{A.~N.}
  \bibnamefont{Musienko}}, \bibinfo{journal}{Atomic Energy}
  \textbf{\bibinfo{volume}{68}}, \bibinfo{pages}{162}
  (\bibinfo{year}{1990}{\natexlab{a}}).

\bibitem[{\citenamefont{Baranenko
  et~al.}(1990{\natexlab{b}})\citenamefont{Baranenko, Fal'kovskii, Kirov,
  Kurnyk, Musienko, and Piontkovskii}}]{Baranenko-1990b}
\bibinfo{author}{\bibfnamefont{V.~I.} \bibnamefont{Baranenko}},
  \bibinfo{author}{\bibfnamefont{L.~N.} \bibnamefont{Fal'kovskii}},
  \bibinfo{author}{\bibfnamefont{V.~S.} \bibnamefont{Kirov}},
  \bibinfo{author}{\bibfnamefont{L.~N.} \bibnamefont{Kurnyk}},
  \bibinfo{author}{\bibfnamefont{A.~N.} \bibnamefont{Musienko}},
  \bibnamefont{and} \bibinfo{author}{\bibfnamefont{A.~I.}
  \bibnamefont{Piontkovskii}}, \bibinfo{journal}{Atomic Energy}
  \textbf{\bibinfo{volume}{68}}, \bibinfo{pages}{342}
  (\bibinfo{year}{1990}{\natexlab{b}}).

\bibitem[{\citenamefont{Yamamoto et~al.}(1976)\citenamefont{Yamamoto,
  Alcauskas, and Crozier}}]{Yamomoto-et_al-1976}
\bibinfo{author}{\bibfnamefont{S.}~\bibnamefont{Yamamoto}},
  \bibinfo{author}{\bibfnamefont{J.~B.} \bibnamefont{Alcauskas}},
  \bibnamefont{and} \bibinfo{author}{\bibfnamefont{T.~E.}
  \bibnamefont{Crozier}}, \bibinfo{journal}{{Journal of Chemical \& Engineering
  Data}} \textbf{\bibinfo{volume}{21}}, \bibinfo{pages}{78}
  (\bibinfo{year}{1976}).

\bibitem[{\citenamefont{Duan and Mao}(2006)}]{Duan-Mao-2006}
\bibinfo{author}{\bibfnamefont{Z.}~\bibnamefont{Duan}} \bibnamefont{and}
  \bibinfo{author}{\bibfnamefont{S.}~\bibnamefont{Mao}},
  \bibinfo{journal}{Geochim.\ Cosmochim.\ Acta}
  \textbf{\bibinfo{volume}{70}}, \bibinfo{pages}{3369} (\bibinfo{year}{2006}).

\bibitem[{\citenamefont{Hildebrand and Scott}(1964)}]{Hildebrand-Scott-1964}
\bibinfo{author}{\bibfnamefont{J.~H.} \bibnamefont{Hildebrand}}
  \bibnamefont{and} \bibinfo{author}{\bibfnamefont{G.~D.} \bibnamefont{Scott}},
  \emph{\bibinfo{title}{The Solubility of Nonelectrolytes}}
  (\bibinfo{publisher}{Dover Publications}, \bibinfo{address}{New York},
  \bibinfo{year}{1964}), \bibinfo{edition}{3rd} ed.

\bibitem[{\citenamefont{Goff}(1957)}]{Goff-1957}
\bibinfo{author}{\bibfnamefont{J.~A.} \bibnamefont{Goff}},
  \bibinfo{journal}{Transactions of the American society of heating and
  ventilating engineers} \textbf{\bibinfo{volume}{70}}, pp.\ \bibinfo{pages}{347} (\bibinfo{year}{1957}).


\end{thebibliography}

\end{document}